\documentstyle[11pt,newpasp,psfig,epsf]{article}

\begin{document}

\title{HST/STIS Spectroscopy of 3C~273}
\author{S. R. Heap, G. M. Williger }
\affil{Code 681, NASA's Goddard Space Flight Center, Greenbelt MD 20771}
\author{R. Dav\'{e}}
\affil{Steward Observatory, University of Arizona}
\author{R. J. Weymann}
\affil{Carnegie Observatories (OCIW)}
\author{E. B. Jenkins, T. M. Tripp}
\affil{Princeton University Observatory}
\begin{abstract}
We present preliminary results on the low-redshift Ly$\alpha$ forest as based
on STIS spectra of 3C~273. A total of 121 intergalactic Ly$\alpha$-absorbing
systems were detected, of which 60 are above the 3.5$\sigma$ completness limit, 
log$N_{\rm HI}\approx$12.3.  
The median line-width parameter, $b=27$ km/s, is similar to that seen
at high redshift. However the distribution of HI column densities has
%$d\mathcal{N}$($N_{\rm HI}$ ) has
a steeper slope, $\beta =2.02 \pm 0.21$, than is seen at high redshift.
Overall, the observed $N_{\rm HI}$--$b$ distribution 
is consistent with that derived from a $\Lambda$CDM hydrodynamic simulation.

\end{abstract}

\section{Introduction}

3C273 is the most closely observed target for studies of the Ly$\alpha$ forest
at low-redshift. As shown in Table 1 (next page), it has 
been observed by FUSE and each of the Hubble spectrographs -- FOS, 
GHRS, and in this past year, the Space Telescope Imaging Spectrograph (STIS). 
It is a cinch for early observation with the Cosmic Origins Spectrograph
(COS) scheduled to be installed on HST in early 2004. 
%(As a COS Co-I, I will make sure of that!)
This interest in 3C273 is for good reason: as the brightest
AGN at a significant redshift (z=0.158) in the sky, it offers
the best opportunity to explore the low-redshift Ly$\alpha$ forest.
This opportunity was not lost on Ray and John Bahcall.
In fact, the first major result of HST came from GHRS and FOS spectroscopy
of 3C~273: the rate of evolution of absorbers, i.e. the rate at which 
absorbers of a given column density disappear from view, slows dramatically 
at $z < 2$ (Morris et al. 1991; Bahcall et al. 1991). The survival of 
HI-absorbing systems to the present epoch, which was predicted by Ray and his 
colleagues (Bechtold, Weymann et al. 1987), is thought to be  
a  consequence of a lower ionizing background at $z < 2$,  reflecting 
a decline in the quasar population  after $z\sim 2$ (Dav\'{e} et al. 1999). 

Cosmological simulations indicate that low--$z$ and high--$z$ absorbing
systems having the same column density belong to different density/structure  
regimes. % in the intergalactic medium (IGM). %Because of the thinning-out
%of the universe due to cosmological expansion and the growth of structure, 
For example, an absorption
feature of a given column density arising in  a filament  in the low-$z$ IGM
would arise in a relative void at high-redshift.  
This puts investigators of the low--$z$ IGM at a disadvantage with respect
to our high-$z$ colleagues. We have to pry out  
weak  absorption features to explore physical systems that would produce saturated
absorbers at high redshift. This is why the STIS spectrum
of 3C~273 is so important. For the first time, we can measure absorbing
systems down to a completeness %at a $4.0\sigma$ detection threshold 
limit of log$N_{\rm HI}\approx$12.3, thereby probing these weak absorbers
that are predicted to be physically equivalent to high-redshift
forest absorbers.

\begin{center}
Table 1: Observational \& archival studies of the Lyman  Forest toward 3C~273
\begin{tabular}{lccrll}
\hline
Inst & Obs/ & $\lambda\lambda$  & Res  & S/N & Reference \\ 
           & Arch &     (\AA )        &  (km/s)    &     &           \\ \hline
\textsc{FOS} & \multicolumn{1}{c}{Obs} & \multicolumn{1}{c}{1150-1600} & $\sim 240$
& \multicolumn{1}{c}{48} & Bahcall et al. 1991 \\ 
\textsc{FOS} & \multicolumn{1}{c}{Arch} & \multicolumn{1}{c}{} &  & 
\multicolumn{1}{c}{} & Bahcall et al. 1993 \\ 
\textsc{GHRS} & \multicolumn{1}{c}{Obs} & \multicolumn{1}{c}{
1175-1452} & $\sim $200   & \multicolumn{1}{c}{30} & Morris et al. 1991 \\ 
\textsc{GHRS} & \multicolumn{1}{c}{Obs} & \multicolumn{1}{c}{
1235-1425} & $>$20   & \multicolumn{1}{c}{9} & Brandt, Heap et al. 1993 \\ 
\textsc{GHRS} & \multicolumn{1}{c}{Arch} & \multicolumn{1}{c}{} &  
& \multicolumn{1}{c}{ } & Morris et al. 1993 \\ 
\textsc{GHRS} & \multicolumn{1}{c}{Arch} & \multicolumn{1}{c}{} & 
& \multicolumn{1}{c}{} & Brandt, Heap et al. 1997 \\ 
\textsc{GHRS} & \multicolumn{1}{c}{Obs} & \multicolumn{1}{c}{
1215-1250} & $>$20 & \multicolumn{1}{c}{9} & Weymann, Rauch et al. 1995 \\ 
\textsc{GHRS} & \multicolumn{1}{c}{Arch} & \multicolumn{1}{c}{ } 
&  & \multicolumn{1}{c}{ } & Grogin \& Geller 1998 \\ 
\textsc{GHRS} & \multicolumn{1}{c}{Arch} & \multicolumn{1}{c}{} & 
& \multicolumn{1}{c}{} & Penton, Stocke et al. 2000 \\ 
\textsc{GHRS} & \multicolumn{1}{c}{Arch} & \multicolumn{1}{c}{} & 
& \multicolumn{1}{c}{} & Penton, Shull et al. 2000 \\ 
\textsc{FUSE} & \multicolumn{1}{c}{Obs} & \multicolumn{1}{c}{900-1190} & 20 & 
\multicolumn{1}{c}{15-35} & Sembach et al. 2001 \\ 
\textsc{STIS} & \multicolumn{1}{c}{Obs} & \multicolumn{1}{c}{1150-1700}& 7  
& \multicolumn{1}{c}{30-44} & Heap et al. 2001 (in prep) \\ \hline
\end{tabular}
\end{center}

\section{Observations}

To observe 3C~273 we used STIS in the E140M mode, which yields
an echelle spectrogram covering the spectral range, 1150--1700 \AA .
The nominal  two-pixel resolution of this mode is 7 km/s, but in our case, 
the effective resolving power was slightly less because we used a ``large'' 
0\farcs{2} aperture and therefore lost some spectral purity. We observed 3C273 
during two visits spaced a few months apart, the first one lasting 2 orbits, the
second, for 5 orbits. The total exposure time was 18,671 sec.
The two visits have the spectrum on  different 
regions of the detector format, which means that
every point in the spectrum was viewed by two different
regions of the detector. This redundancy made it easier to identify
and account for detector blemishes and fixed-pattern noise.

\begin{figure}[ht]
\centerline{\vbox{
\psfig{figure=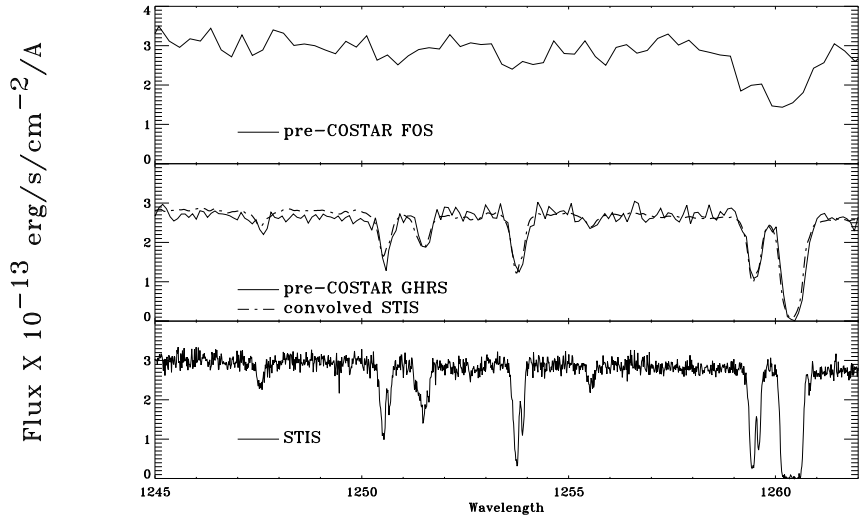,height=3in}}}
\caption[]{FOS (top), GHRS (middle), and STIS (bottom) spectra of 3C~273. 
The middle panel also shows our STIS spectrum convolved with the line-spread
function of the pre-COSTAR GHRS (Heap et al. 1995). The features at 1250.5, 
1253.8, 1259.5, and 1260.4 \AA\ are interstellar.}
\end{figure}

We reduced the data using software developed at Goddard by 
the STIS Instrument Definition Team (Lindler 1998). Individual orders
of the echellogram were extracted and calibrated, and corrections 
for scattered light were then applied. %To create the final, 1-D spectrum, 
We combined the 7 orbits of data for each spectral order,
with weighting based on a pixel's signal-to-noise.
We then coadded the spectral orders to form a single, continuous spectrum.
The S/N per pixel rises from $\sim 20$ at 1230 \AA\ to $\sim 29$ at
1380 \AA, although there is significant variation in S/N within each order.
Equivalently, the S/N per 7-km/s resolution element is 30 -- 44.  % /stis/may/re.jou

Figure 1 compares a segment of the final STIS spectrum with those from
the FOS and GHRS. It demonstrates the marked improvement in S/N and 
resolution over previous spectrographs. Actually, the comparison with GHRS 
(middle panel) is a little unfair, since the only GHRS spectra of 3C~273 
were taken before the COSTAR corrective optics were deployed. Nevertheless,
it shows that pre-COSTAR GHRS spectra couldn't resolve features into multiple
components as STIS can. In fact, STIS can resolve virtually all 
Ly$\alpha$ absorbers.

\section{Simulations}

In parallel with STIS data analysis, we generated 20 artificial STIS spectra 
drawn from a simulation of a $\Lambda$CDM universe (see Dav\'e et al. 1999 for
simulation parameters).
By comparing the statistical properties of the observed and artificial spectra,
we can in principle (i) constrain the parameters of the $\Lambda$CDM  model and 
(ii) infer physical conditions in the local intergalactic medium.

It is crucial that the simulations be faithful reproductions of STIS data 
if they are to be of any value. We therefore took great pains
to account for all the characteristics and foibles of STIS data. 
Figure 2 illustrates some of the steps to compute the simulated STIS
spectra. First, we took a normalized simulated spectrum (top panel) and put
it on an absolute flux scale. We then constructed the corresponding STIS
echellograms (one for each visit) complete with scattered light, 
noise as appropriate, and interstellar lines added in. We next reduced the 
echellograms in exactly the same way as for a STIS observation and 
estimated the continuum level (middle panel). The bottom panel shows
the final, normalized STIS spectrum ready for measurement with AutoVP 
(Dav\'{e} et al. 1997). 

\begin{figure}[ht]
\centerline{\vbox{
\psfig{figure=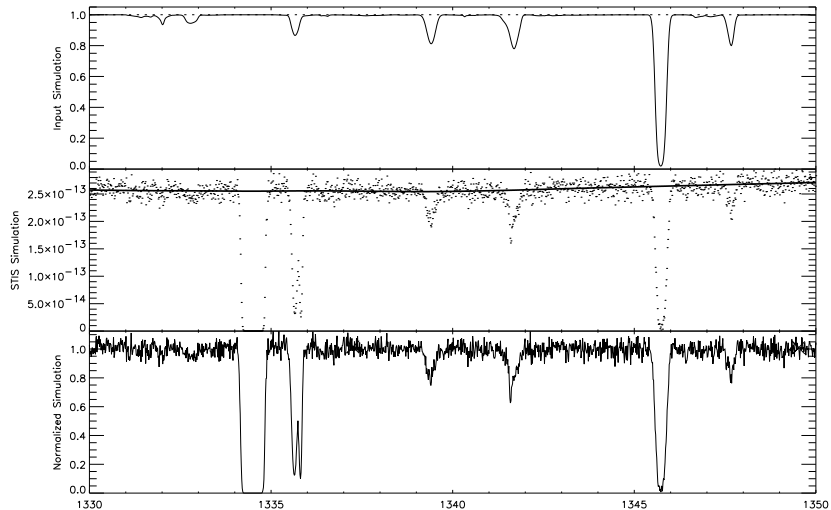,height=3in} }}
\caption[]{Portion of a simulated spectrum. The three panels show the
original, normalized simulated spectrum (top), the STIS simulation with 
estimated continuum (middle), and the final, normalized STIS simulated spectrum.
The strong features near 1335\AA\ are interstellar C\textsc{ii} 
lines that has been added to the artificial spectrum. }
\end{figure}

\section{Measurements}

The preliminary results 
presented here are based on Dav\'{e}'s method for defining the continuum
using an automated median filter, then using AutoVP to 
detect absorption features and determine their
column densities ($N_{\rm HI}$) and Doppler parameters ($b$). 
Independently, GW and SRH also  
defined the continuum level (using manual procedures), 
identified absorption systems, and measured line parameters.  We  are now 
in the process of comparing our results in detail, but one early finding
is that lines that
are unsaturated and isolated (which constitute most of the intergalactic
absorbers detected) have line parameters that are consistent within 
errors between these three independent estimates.

RD has recently modified AutoVP in an attempt
to identify more broad, weak absorbers that may arise in 
shock-heated intergalactic gas (e.g. Tripp, Savage \& Jenkins 2000).  
In particular, AutoVP now uses a variable-width Gaussian detection window 
rather than a narrow, fixed-width boxcar window, as in previous studies 
such as Dav\'e \& Tripp (2001).  We are in the process of extensively testing 
the effects of this new algorithm, but preliminary results seem to indicate 
that significantly more broad, weak absorbers are now detected.

$b$-$N_{\rm HI}$ distributions are one
way to gain information about the ``effective equation of state'' of
the IGM at low-redshift. 
Since the narrowest absorbers are presumably purely thermally broadened,
the lower envelope of the  $b$-$N_{\rm HI}$ distribution gives a measure
of the temperature of Ly$\alpha$ absorbing gas that has not been shock-heated 
(Hui \& Gnedin 1997).

How much faith should we put in our $b$-$N_{\rm HI}$ measurements? The FUSE Ly$\beta$ 
survey should give us pause. Shull et al. (2000; this conference) found that 
Doppler widths of \textit{saturated} lines (log$N_{\rm HI}\ga 13.7$) derived from 
Ly$\beta$/Ly$\alpha$ equivalent-width ratios were systematically lower than from Ly$\alpha$ 
profile analysis of GHRS spectra; consequently, the FUSE column densities are higher. This
appears to be the case with 3C~273 as well. Penton et al. (2000a,b) derived
high Doppler widths ($b \sim 60-80$ km/s) for 13 Ly$\alpha$ lines detected in pre-COSTAR GHRS
spectra of 3C~273, whereas Sembach et al. (2001) 
got much lower Doppler widths from Ly$\beta$/Ly$\alpha$ ratios in the
FUSE spectrum for the 8 strong absorbers in common. 

Fortunately, the situation is not as serious as would appear at first glance -- at 
least for 3C~273.  Our $b$ estimates from STIS spectra generally agree within
the quoted $1\sigma$ errors with the COG estimates from FUSE. The only
exception is the strong, saturated absorber at 1222\AA\ (which is
not included in our study). The spurious GHRS results were caused by 
a failure to account for the blurring of pre-COSTAR spectra (Fig. 1).
They are not relevant to our study since most (97\%) of the 
Ly$\alpha$ lines in the STIS spectrum are unsaturated and should therefore
have reliable line parameters.

%2. In our b-param comparison with GHRS/FUSE, we should
%mention that line parameters are MUCH less reliable when
%the absorber starts to become saturated.  In unsaturated 
%cases, virtually all fiting programs give values that
%agree within errors.  So in the cases of the higher
%column lines examined by preious groups, it is not
%surprising that there are large variations between 
%methods.  Most of our lines, however, are unsaturated,
%and therefore should have reliable line parameters.

\begin{figure}[h]
\centerline{\vbox{
\psfig{figure=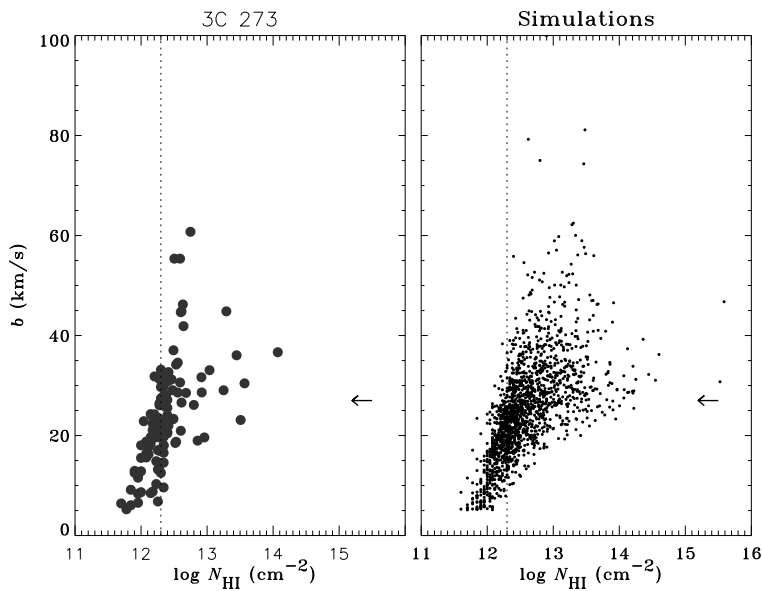,height=3in }  }}
\caption[]{ $N_{\rm HI}$-$b$ for the 3C~273 observations vs. 
simulations. The vertical dotted line indicates the estimated 
completeness limit (log$N_{\rm HI}$=12.3); the horizontal arrow, the measured median Doppler 
width of lines above the completeness limit (med. $b$=27 km/s for both observations and simulations). }
\end{figure}

\section{Results}

Figure 3 shows our measured $b$-$N_{\rm HI}$ distributions for 3C~273 and the
simulations. Both distributions show a strong correlation between column
density and linewidth.
The median Doppler width in the simulations, where
we have better statistics, increases quickly from about 20 km/s at $N_{\rm HI}$=12.3 
to 30 km/s at $N_{\rm HI}$=13.0, then begins to level off for $N_{\rm HI}\ga $14.0.  The levelling off
suggests that these high column density absorbers are probing gas  
associated with galaxies, and is not heated by the metagalactic ultraviolet flux.

The median Doppler width found for 3C273 is significantly higher than what Dav\'e \& Tripp 
(2001) measured %, $b\approx 21$~km/s, 
from STIS spectra of PG0953 and H1821 (Table 2).
This difference likely reflects the new line-detection algorithm in AutoVP described
earlier.  In fact, a reanalysis of the Dav\'e \& Tripp data set with the
new AutoVP also finds a higher median $b\approx 26$~km/s.  This example illustrates
how specific algorithmic choices can have a significant impact on 
line parameters and resulting interpretations, a fact that is well-known 
in high-redshift forest
studies (motivating, for instance, the use of flux-based statistics).

\begin{center}
Table 2: Comparison with other studies

\begin{tabular}{lcccc}
\hline
Reference  & log \textit{N}$_{\mathrm{HI}}$ range & \textit{{z}} &  $ \beta $ & med. \textit{b} \\ \hline
{Kim et al. (2001)}    &   12.5-14.0  & {2.1}              & 1.38$\pm 0.08$     & 25  \\ 
{Kim et al. (2001)}    &   12.5-14.0  & {1.6}              & {1.72$\pm 0.16$}   & 28  \\ 
{Shull (this conf.). } &   12.3-14.0  & {$<0.1$}           & {1.82$\pm 0.10$}   & 25* \\ 
{Dav\'{e} \& Tripp (2001)}&13.0-14.5  & {0.17}             & {2.04$\pm 0.23$}   & 21  \\ 
{This paper}           & 12.3-14.0    & {$<0.1$}           & {2.02$\pm 0.21 $}  & 27  \\ \hline
\end{tabular}

 *value assumed for COG estimation of log \textit{N}$_{\mathrm{HI}}$
\end{center}

The top panel of Figure 4 shows the number density distribution, 
$\mathcal{N}$($N_{\rm HI}$)$=A N_{\rm HI}^{-\beta}$, 
for both 3C~273 and the simulations.  The turnover in this distribution
at low column densities allows a rough estimate of the completeness limit
of lines, and is seen to be roughly $N_{\rm HI}\approx 10^{12.3}$~cm$^{-2}$;
we are working on a more thorough analysis of completeness in this sample.
The observed slope for 3C~273, $\beta \approx 2.0$
is slightly steeper than that estimated by Penton et al. (2000; c.f. Shull, this conference) 
but is nearly exactly the same as that found by Dav\'{e} and Tripp (Table 2).

 \begin{figure}[h]
\centerline{\vbox{
\psfig{figure=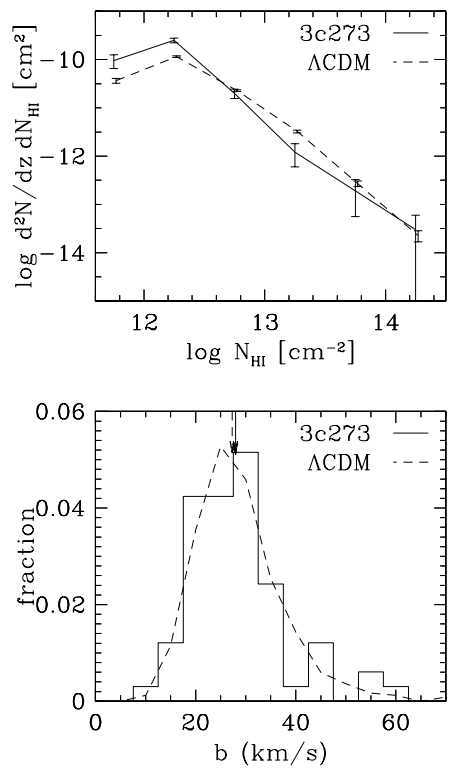,height=4.7in} }} 
\caption[]{ Column density distributions (top) and Doppler-width distributions
(bottom). In both panels, the observations are indicated by the solid line,
and the simulations, by the dashed line.}
\end{figure}

Both our observations and simulations have slopes  
($\beta_{obs} = 2.02 \pm 0.21$, $\beta_{sim} = 1.90 \pm 0.04$) that are 
significantly steeper than at $z>2$ (see Table 2).
%, where $\beta_{obs} \sim 1.5$ was measured 
At higher column densities, the $\mathcal{N}$($N_{\rm HI}$) distribution 
flattens somewhat ($\beta = 1.42 \pm 0.16$ according to Shull, this conference)
and comes into basic agreement with the high-redshift slope (Kim et al. 2001).
The steeper slope at low column densities indicates that higher column-density 
lines are evolving away faster than are lower column-density ones, in 
agreement with findings from the HST Quasar Absorption-Line Key Project (Weymann et al. 1998).

The agreement in amplitude indicates that the ionizing background assumed in the simulations
is close to correct (within the framework of this cosmology and simulation; 
see discussion in Dav\'e \& Tripp);   it seems to suggest
the intensity predicted by Haardt \& Madau (1996) is roughly
correct (within a factor of a few), as suggested by the analysis
of Dav\'e \& Tripp.

\end{document}